\documentclass[11pt,a4paper]{article}

\usepackage[margin=1in]{geometry}

\usepackage[T1]{fontenc}
\usepackage[utf8]{inputenc}
\usepackage{newtxtext,newtxmath}
\usepackage{amsmath}

\usepackage{graphicx}
\usepackage{booktabs}
\usepackage{multirow}
\usepackage{array}

\usepackage{caption}
\usepackage{subcaption}

\usepackage[numbers,sort&compress]{natbib}

\usepackage[hidelinks]{hyperref}
\usepackage[nameinlink,noabbrev]{cleveref}

\usepackage{lineno}

\usepackage{authblk}

\usepackage{xurl}

\usepackage{microtype}

\title{GraphIDyOM: A graph-native Python reimplementation of IDyOM for musical expectation modelling}

\author[1]{Lluc Bono Rosselló}
\affil[1]{Institute for Interdisciplinary Studies on Artificial Intelligence (IRIDIA), 

Université Libre de Bruxelles, Brussels, Belgium

\texttt{Lluc.Bono.Rossello@ulb.be}}

\begin{document}
\date{}
\maketitle

\begin{abstract}
The Information Dynamics of Music model (IDyOM) has played a central role in computational accounts of musical expectation by providing event-by-event estimates of uncertainty and surprise from symbolic musical sequences. However, its reference implementation is difficult to integrate with contemporary Python workflows, and its internal memory structures are not easily accessible for inspection or modification. We introduce GraphIDyOM, a graph-native Python reimplementation of IDyOM that represents long-term and short-term predictive memories as explicit graph objects while preserving the model's variable-order, multiple-viewpoint architecture. GraphIDyOM returns event-wise information content and entropy, exposes internal memory structures for analysis and export, and supports access through a local server. We validate the implementation against the original Lisp IDyOM across single, projected, and multiple-viewpoint configurations, and benchmark its coverage and computational performance against a recent reimplementation. We then demonstrate how the explicit memory representation supports network analysis of learned memories, projection of expectation values onto musical networks, recency-sensitive memory retrieval, and interactive applications. GraphIDyOM therefore provides both a faithful and accessible reimplementation of a widely used model and a platform for studying musical expectation through memory, topology, and interaction.
\end{abstract}

\section{Introduction}

Musical listening unfolds under expectation. A melody is not experienced only
as a sequence of events that have occurred, but also as a changing field of
possible continuations shaped by prior exposure, local context, and learned
musical conventions \citep{Saffran1999ToneSequences,huron2008sweet,temperley2007music}. Computational models of musical expectation formalise this process by estimating, at each
point in a sequence, a probability distribution over what may happen next. From
this distribution, two information-theoretic quantities are central: entropy,
which measures predictive uncertainty before an event occurs, and information
content, which measures the surprise of the event that is actually observed
\citep{pearce2012auditory,pearce2018statistical}.

The Information Dynamics of Music model (IDyOM) is one of the most influential
frameworks for this kind of analysis
\citep{pearce2012auditory,pearce2018statistical}. It models musical expectation
from symbolic representations of musical events and relations, combining
regularities learned from prior exposure with those emerging from the
currently unfolding sequence. Its multiple-viewpoint and variable-order
architecture allows expectations to depend on representation, context depth,
stylistic exposure, and local adaptation
\citep{conklin1995multiple,cleary1984data}. This listener-relative approach has
made the model useful across cognitive neuroscience, music psychology, and
empirical aesthetics. IDyOM-derived uncertainty and surprise have been related
to neural responses \citep{di_liberto_cortical_2020,quiroga-martinez_decomposing_2020,kern_cortical_2022,koelsch2019predictive,bianco_neural_2024}, musical
imagery \citep{marion_music_2021}, perceived complexity
\citep{sauve_information-theoretic_2019}, liking
\citep{gold2019predictability,mas-herrero_predictive_2025,
cheung2019uncertainty,bianco2019music}, and sensitivity to learned musical
regularities \citep{bianco2020long}. These applications depend not only on the
final numerical outputs of the model, but also on the ability to select
appropriate viewpoints, reproduce prediction settings, inspect the evidence
supporting particular predictions, and adapt the model to specific analytical
or cognitive assumptions.

The original IDyOM Lisp implementation remains the reference system, but it can
be difficult to install, extend, and integrate with contemporary Python-based
music-analysis workflows. Py2LispIDyOM improves access by wrapping the Lisp
system from Python, but the predictive machinery still runs inside the
underlying Lisp environment and exposes mainly the resulting outputs
\citep{guan2022py2lispidyom}. IDyOMpy provides a useful native Python baseline,
but it implements a smaller modelling surface than the reference architecture
and does not expose all model configuration choices needed for direct parity
\citep{marion2025idyompy}. There is therefore a gap between needs that are often required together: faithful reproduction of the reference model, ordinary
Python accessibility, and direct inspection of the internal memory structures
that generate its predictions.

Making those internal structures explicit is important not only for software
accessibility and reproducibility, but also for addressing a broader
methodological gap in how musical structure and musical expectation are
currently studied. Musical structure can be characterised independently of a
listener model through regularities present in the musical artifact itself.
Statistical
\citep{zanette2006zipf,perotti2020zipf,serra-peralta2021heaps,
nelias2024stochastic}, geometrical
\citep{tymoczko2006geometry,tymoczko2011geometry}, and topological
\citep{liu_tse_small_2009,ferretti2017,frottier2022harmonic,
gomez_lorimer_stoop_2014,kulkarni2024information,rossello2025network,
nardelli_tonal_2021,di2025decoding} approaches have described music in terms
of transition frequencies, recurrence, long-range organisation, tonal or
harmonic structure, vocabulary growth, and relations between musical states.
Such approaches characterise properties of the music as an organised object.

Predictive models such as IDyOM provide a complementary, listener-relative
description: the same event or transition is evaluated according to
regularities acquired through prior exposure and the context accumulated while
listening. This distinction is particularly relevant for complexity, since
structural properties of a piece need not coincide with the complexity
attributed to it by a listener model; IDyOM-derived entropy and information
content have been used to account for perceived musical complexity
\citep{sauve_information-theoretic_2019}. Musical organisation can therefore
be described both through structure intrinsic to the artifact and through the
expectations that a particular model brings to it.

These perspectives are complementary but are typically represented and
analysed separately. Artifact-centred approaches describe the organisation of
musical events and relations, whereas predictive models return
listener-relative quantities along an unfolding sequence. Graph representations provide a common interface between these two levels. The
transitions of a musical sequence can be represented as a network whose
topology reflects the organisation of the artifact, while the predictive
memories that generate IDyOM expectations can be represented through related
graphs encoding learned context--continuation statistics. This common representation makes it possible to relate
artifact-level organisation directly to listener-relative prediction: not only
asking how a piece is structured or how surprising it is to a particular
model, but where and how learned expectations are organised over the structure
through which they arise.

The same explicit access to predictive memory also makes it possible to test
alternative assumptions about how stored evidence remains available over time.
For example, recency-sensitive PPM models allow older observations to
contribute less strongly to prediction than more recent ones
\citep{harrison2020ppm,bianco2020long}. An IDyOM-compatible framework in which
memory weights can be modified directly therefore provides a way to study such
assumptions without replacing the surrounding prediction architecture.

Explicit and queryable access to the predictive model also broadens how
expectation can be used experimentally and creatively. Rather than computing
entropy and information content only after a sequence has been generated,
successive predictions can be queried while the sequence is being constructed.
This allows musical stimuli to be selected or adapted according to target
predictive properties, and enables interactive systems to use next-event
probabilities as interpretable constraints or control signals during musical
interaction.

This paper introduces GraphIDyOM, a Python reimplementation and extension
framework for IDyOM that represents long-term and short-term variable-order
predictive memories as explicit graphs of contexts and weighted continuations.
The formulation preserves the count-based prediction machinery of the reference
model while making the internal structures that support prediction directly
accessible for analysis and modification.

We first evaluate numerical correspondence with the original Lisp
implementation and benchmark the resulting Python framework. We then show how
this graph-native representation enables network analysis of learned memories
and expectation-annotated musical structures, recency-sensitive memory
retrieval, and external access to trained models for experimental and
interactive applications. Together, these extensions position GraphIDyOM as a
faithful and accessible reimplementation of IDyOM that also provides a common
representation for analysing, modifying, and interactively querying musical
expectation.

\section{GraphIDyOM}

IDyOM estimates musical expectation from statistical regularities in symbolic
musical sequences \citep{conklin1995multiple,pearce2012auditory,
pearce2018statistical}. Musical events and relations between them are encoded
through \emph{viewpoints}, such as chromatic pitch, melodic interval, duration,
or inter-onset interval \citep{conklin1995multiple}. Predictions are informed
by two sources of statistical memory: a long-term model (LTM), learned from a
training corpus, and a short-term model (STM), updated as the current sequence
unfolds. Within each memory, Prediction by Partial Matching (PPM) combines
evidence across multiple Markov orders, allowing longer and more specific
contexts to contribute when available while progressively shorter contexts
provide fallback predictions \citep{cleary1984data}.

Functionally, the model maps the current history \(h_t\) onto a probability
distribution \(p(\cdot \mid h_t)\) over possible continuations. Two quantities
derived from this distribution are central throughout this work. Predictive
uncertainty is quantified by entropy,

\begin{equation}
H_t = -\sum_i p_i \log_2 p_i,
\end{equation}

while the surprise associated with the event that actually occurs is quantified
by its information content,

\begin{equation}
IC(e_t) = -\log_2 p(e_t \mid h_t).
\end{equation}

GraphIDyOM reimplements this architecture in Python while making its predictive
memories explicit as directed graph objects. IDyOM predictions ultimately
depend on counts of contexts and their observed continuations. In the
graph-native implementation, these structures are represented directly: for
each Markov order, nodes encode contexts and weighted directed edges encode
observed continuations. LTM graphs are learned from the training corpus,
whereas STM graphs are constructed and updated as the target sequence unfolds.
The graph representation therefore does not replace the variable-order
statistical machinery of IDyOM; it provides an explicit computational
representation of the memory on which that machinery operates.

The implementation preserves the principal components required for IDyOM-style
prediction, including direct and derived viewpoints, projection between
representational spaces, multiple-viewpoint prediction, variable-order PPM,
and separate LTM and STM models. Because the graph objects used for prediction
are directly accessible, the same memories can also be inspected, exported,
analysed, or selectively transformed. This explicit representation provides
the basis for the network analyses, alternative memory mechanisms, and
interactive applications introduced in Section~\ref{sec:extensions}.

\subsection{Prediction architecture and processing pipeline}

GraphIDyOM follows the same prediction sequence as IDyOM, from the symbolic
encoding of a musical event to a final probability distribution over possible
continuations. The process begins by representing the musical input through
one or more \emph{viewpoints}, then matching the resulting symbolic history
against memories stored at different Markov orders. Evidence is combined first
across orders and then across long- and short-term memories to obtain the final
next-event distribution.

In IDyOM, viewpoints map musical events or relations between events onto
symbolic sequences that can be modelled statistically
\citep{conklin1995multiple}. The implementation preserves this distinction
between the musical surface and the symbolic representation used for
prediction. MIDI input is first parsed into an ordered sequence of note events,
after which viewpoint functions encode event attributes, relations between
events, or linked combinations of several attributes.

The current implementation includes pitch- and timing-related basic viewpoints,
derived representations such as interval, contour, and temporal ratios, and
linked viewpoints that combine several attributes into a single symbolic state.

As in IDyOM, the model distinguishes the \emph{source viewpoint} in which
contexts are learned from the \emph{target viewpoint} over which the final
prediction is evaluated. When source and target are equivalent, prediction is
direct. When they differ, predictions made in the source space are projected
onto the target space. For example, if the current pitch is \(60\), probability
assigned by an interval viewpoint to a continuation of \(+2\) can be projected onto
candidate pitch \(62\).

Multiple viewpoints can be combined in two ways. Several attributes may form a
joint target, such as pitch and duration predicted together, or several source
viewpoints may contribute evidence to a common target. In the latter case,
their distributions are first projected into the same target space and then
merged. Entropy-weighted merging gives greater influence to more concentrated
component distributions. These operations reproduce the source--target and
multiple-viewpoint logic of the reference architecture.

Once a source-viewpoint sequence has been obtained, prediction proceeds by
matching its recent history against contexts learned at different Markov
orders. For a source sequence \(x_1,\ldots,x_T\), a model of order \(n\)
conditions prediction on the context

\[
c_t^{(n)}=(x_{t-n},\ldots,x_{t-1}).
\]

GraphIDyOM represents each order as a directed graph \(G^{(n)}\), whose nodes
are \(n\)-gram context windows and whose weighted edges record observed
transitions between successive contexts. Prediction at a particular order
therefore consists of locating the node corresponding to the current context
and following its outgoing edges. Each edge represents an observed continuation
and points to a successor \(n\)-gram whose final element encodes the candidate
next symbol; the edge weight gives the corresponding continuation count. Normalising these
counts produces an order-specific distribution \(p_n(s\mid h_t)\) over
possible next symbols.

\begin{figure}[htbp!]
    \centering
    \includegraphics[width=0.8\textwidth]{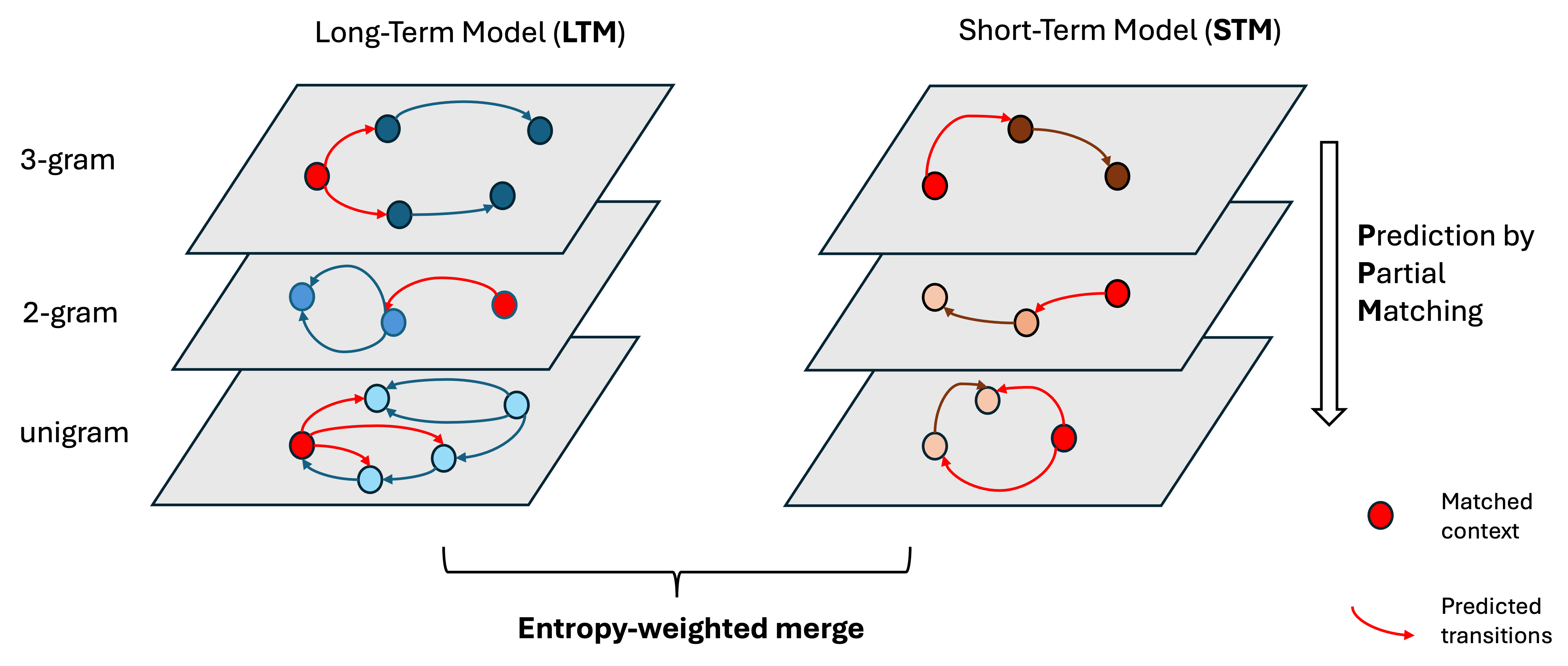}
    \caption[GraphIDyOM schematic]{
    \textbf{GraphIDyOM schematic.}
    Symbolic input is queried against order-specific long-term (LTM) and
    short-term (STM) memory graphs. At each order, the current context is
    matched to a node and outgoing weighted edges define possible
    continuations. PPM combines evidence across progressively shorter
    contexts within each memory, after which the LTM and STM distributions
    are merged to produce the final next-event distribution from which
    entropy and information content are computed.
    }
    \label{fig:graphidyom_layers}
\end{figure}

The resulting object is an \(n\)-gram memory graph rather than a reduction of
the musical sequence to first-order transitions. Higher-order graphs encode
progressively longer histories, so the same current event may lead to different
predictions depending on the sequence that preceded it. As shown in
Figure~\ref{fig:graphidyom_layers}, the current history is queried across these
order-specific graphs. If a longer context is unavailable or does not account
for all possible continuations, prediction falls back to shorter histories.

The next step is to combine the evidence available at different orders.
Prediction by Partial Matching (PPM) assigns probability first from the
longest available context and passes residual probability mass, through an
escape mechanism, to progressively shorter contexts
\citep{cleary1984data}. In this way, specific higher-order evidence is used
when available while shorter contexts provide increasingly general fallback
predictions. In compact form,

\begin{equation}
P_{\mathrm{PPM}}(s \mid h_t) =
\sum_{n=0}^{K} \alpha_n p_n(s \mid h_t)
+ \alpha_{-1}p_{-1}(s),
\end{equation}

where \(K\) is the maximum order and \(\alpha_n\) denotes the probability mass
reaching each order after successive escapes. The residual term
\(\alpha_{-1}p_{-1}(s)\) represents the final fallback distribution.

The implementation supports the standard IDyOM PPM escape methods
\texttt{a}, \texttt{b}, \texttt{c}, \texttt{d}, and \texttt{x}, together
with exclusion and update-exclusion options. These choices determine how
continuation counts and escape mass are handled across orders and are required
to reproduce the distinct LTM and STM configurations of the reference
implementation.

This variable-order prediction process is performed separately for the
long-term and short-term memories. LTM and STM use the same graph
representation and PPM machinery but differ in how their evidence is acquired.
The LTM is learned from a corpus and remains fixed while a target piece is
processed. The STM is reset at the beginning of each target piece and updated
after every observed event, allowing the model to adapt to regularities that
emerge within the current sequence. Either memory can be evaluated
independently, or both can contribute to prediction.

When both memories are active, the LTM and STM each produce a probability
distribution over possible next events. These distributions are then combined
into a single prediction. Rather than weighting both memories equally, IDyOM
gives greater influence to the distribution that is more confident, that is,
the one whose probability mass is more concentrated over a smaller set of
candidates.

This confidence is estimated from the entropy of each component distribution.
For a distribution \(p_m\), its entropy is normalised by the maximum entropy
possible over its support,

\[
r_m =
\frac{H(p_m)}
{\log_2 |\mathrm{supp}(p_m)|},
\]

so that \(r_m\) expresses how uncertain that distribution is relative to a
uniform distribution over the same set of candidates. Lower values therefore
indicate a more concentrated, and hence more confident, prediction. These
relative entropies are converted into normalised combination weights,

\begin{equation}
w_m =
\frac{(r_m+\delta)^{-b}}
{\sum_j (r_j+\delta)^{-b}},
\end{equation}

where \(b\) controls how strongly differences in uncertainty affect the merge
and \(\delta\) prevents numerical problems when entropy is very small. The
weights sum to one, with lower-entropy component distributions receiving
greater influence.

For the geometric combination used in the Lisp-aligned validation
(see Section~\ref{sec:validation}), following
\citet{pearce2004methods}, the component distributions are combined as

\begin{equation}
P(s) =
\frac{\prod_m p_m(s)^{w_m}}
{\sum_{s'} \prod_m p_m(s')^{w_m}},
\end{equation}

where each candidate probability is combined across the active memories
according to their confidence weights and the result is renormalised to form a
valid probability distribution.

The resulting \(P(s\mid h_t)\) is the final distribution over possible next
events in the target viewpoint. Its entropy quantifies uncertainty before the
next event is observed, while the probability assigned to the event that
actually occurs determines its information content.

\subsection{Implementation, interfaces, and outputs}

GraphIDyOM is implemented as a Python framework centered on a core prediction
model responsible for training, prediction, STM updating, PPM combination,
viewpoint projection, LTM/STM merging, and model persistence. Viewpoint
encoding, memory transformations, graph analysis, and application-facing
interfaces are provided as additional components. Optional extensions operate
on the same underlying model and memory objects and do not alter the default
prediction pipeline unless explicitly enabled.

Models can be trained, saved, loaded, and queried directly through the Python
API. Command-line and lightweight local-server interfaces provide additional
access to the same prediction machinery. In particular, the server can keep a
trained model loaded across successive requests, allowing external
applications to submit an unfolding musical context and obtain predictions
without reconstructing or embedding the model. This interface provides the
basis for the experimental and interactive applications discussed in
Section~\ref{sec:external_access}.

At the event level, the framework returns the probability assigned to the
observed event, information content, predictive entropy, and the complete
distribution over candidate continuations. Optional diagnostic traces expose
intermediate quantities including order-specific LTM and STM predictions,
projected distributions, merge weights, and candidate continuations. These
outputs make it possible to inspect the contributions of different context
depths, viewpoints, and memory sources.

The underlying predictive memories remain directly accessible as graph objects
and can be exported in \texttt{GraphML} format with context labels and node and
edge weights preserved. The exported representations can therefore be examined
using network-analysis libraries such as NetworkX
\citep{hagberg_exploring_2008} or visualised using software such as Gephi
\citep{bastian_gephi_2009}.

\section{Validation and benchmarking}
\label{sec:validation}

We evaluated the default GraphIDyOM implementation against the original Lisp IDyOM using 185 monophonic Bach chorale melodies from the corpus used in the original IDyOM work \citep{pearce2005construction}, the same data source used in the IDyOMpy evaluation \citep{marion2025idyompy}.
The validation
used five-fold cross-validation with seed 0. In each fold, the LTM was trained
on four fifths of the corpus and evaluated on the held-out fifth, yielding 9227
event-level comparisons with Lisp IDyOM per configuration. Both implementations
used LTM and STM components, a maximum Markov order of 5, LTM escape method
\texttt{c}, STM escape method \texttt{x}, LTM update exclusion disabled, STM
update exclusion enabled, and geometric entropy-weighted LTM/STM merging with
bias 7.0 and no entropy-weight offset.

The validation covers several forms of representation and prediction supported
by the IDyOM architecture. Direct configurations include chromatic pitch with
octave (\texttt{cpitch}), note duration (\texttt{dur}), and basic inter-onset
interval (\texttt{bioi}). Other configurations use derived viewpoints:
\texttt{cpint} represents signed melodic interval, \texttt{cpint-size} its
unsigned magnitude, \texttt{contour} the direction of melodic motion, and
\texttt{cpcint} a pitch-class interval representation. Temporal derivatives
include relative BIOI (\texttt{bioi-ratio}) and BIOI contour. For derived
viewpoints whose source and target differ, such as interval-to-pitch
prediction, the source distribution is projected into the target space before
evaluation.

We additionally evaluated two multiple-viewpoint configurations. In the
multi-joint condition, pitch and BIOI form a compound target and are predicted
together. In the multi-source condition, pitch and interval provide separate
sources of evidence for a common pitch target. Together, these tests cover
direct, projected, joint-target, and shared-target prediction.

Where IDyOMpy could express an approximately comparable configuration, we also
evaluated it using \texttt{use\_original\_PPM=True}. Because IDyOMpy does not
expose the same escape-method, update-exclusion, target-projection, or full
multiple-viewpoint controls, configurations that could not be matched are
marked with dashes. Table~\ref{tab:bach-pearce-reimplementation-validation}
reports event-wise agreement with Lisp IDyOM, treated as the reference
implementation. For the two Python reimplementations, we report the mean absolute
difference in information content (\(\Delta IC\), in bits) and the Pearson
correlation \(r\) with the corresponding Lisp information-content trace.

\begin{table}[htbp!]
\centering
\footnotesize
\setlength{\tabcolsep}{3pt}
\caption{Bach validation and benchmarking against Lisp IDyOM. Each row
reports event-wise information-content agreement with Lisp IDyOM.
\(\Delta IC\) is the mean absolute IC difference in bits and \(r\) is the
Pearson correlation with the Lisp IC trace. Dashes indicate configurations
that the tested IDyOMpy implementation cannot express.}
\label{tab:bach-pearce-reimplementation-validation}
\begin{tabular}{@{}p{2.5cm}p{1.4cm}p{2.1cm}p{1.8cm}rcccc@{}}
\hline
Configuration & Type & Source vp. & Target vp. & Events &
\multicolumn{2}{c}{GraphIDyOM} & \multicolumn{2}{c}{IDyOMpy} \\
\cline{6-9}
 & & & & & $\Delta IC$ & $r$ & $\Delta IC$ & $r$ \\
\hline
Pitch+octave & Single & \texttt{cpitch} & \texttt{cpitch} & 9227 & 0.0011 & 0.9997 & 1.4085 & 0.7192 \\
Pitch class & Single & \texttt{cpitch-class} & \texttt{cpitch} & 9227 & 0.0010 & 0.9997 & -- & -- \\
BIOI(length) & Single & \texttt{bioi} & \texttt{bioi} & 9227 & 0.0010 & 0.9999 & 0.7064 & 0.6066 \\
Duration & Single & \texttt{dur} & \texttt{dur} & 9227 & 0.0013 & 0.9999 & -- & -- \\
Interval & Single & \texttt{cpint} & \texttt{cpitch} & 9227 & 0.0008 & 0.9998 & -- & -- \\
Interval size & Single & \texttt{cpint-size} & \texttt{cpitch} & 9227 & 0.0009 & 0.9999 & -- & -- \\
Contour & Single & \texttt{contour} & \texttt{cpitch} & 9227 & 0.0022 & 0.9997 & -- & -- \\
CPCINT & Single & \texttt{cpcint} & \texttt{cpitch} & 9227 & 0.0007 & 0.9998 & -- & -- \\
BIOI ratio & Single & \texttt{bioi-ratio} & \texttt{bioi} & 9227 & 0.0018 & 0.9997 & -- & -- \\
BIOI contour & Single & \texttt{bioi-contour} & \texttt{bioi} & 9227 & 0.0012 & 0.9999 & -- & -- \\
Pitch+length & Multi-joint & \texttt{cpitch+bioi} & \texttt{cpitch+bioi} & 9227 & 0.0021 & 0.9999 & 1.8703 & 0.6916 \\
Pitch+interval to pitch & Multi-source & \texttt{cpitch+cpint} & \texttt{cpitch} & 9227 & 0.0010 & 0.9998 & -- & -- \\
\hline
\end{tabular}

\vspace{0.3em}
\begin{minipage}{0.96\textwidth}
\footnotesize
\emph{Note.} Events report the Lisp--GraphIDyOM comparison count; IDyOMpy
length-like rows use the closest aligned subset where required. Lisp IDyOM is
the reference system, so its own \(\Delta IC\) is 0 and \(r=1\) by definition.
The multi-joint row evaluates a joint pitch-and-BIOI target; the multi-source
row evaluates a common pitch target predicted from pitch and interval sources.
\end{minipage}
\end{table}

GraphIDyOM closely reproduced Lisp IDyOM under matched parser, viewpoint, PPM,
update-exclusion, and LTM/STM merge settings. Across direct, projected,
multi-joint, and multi-source configurations, mean absolute
information-content differences remained below 0.003 bits and correlations
with the Lisp traces were at least \(r=.9997\). This agreement therefore
extends beyond direct prediction to cases requiring source-to-target
projection and multiple-viewpoint integration. IDyOMpy could express only a
subset of these configurations and showed substantially larger deviations
from the Lisp reference in the directly comparable cases.

\begin{figure}[htbp!]
    \centering
    \includegraphics[width=0.7\textwidth]{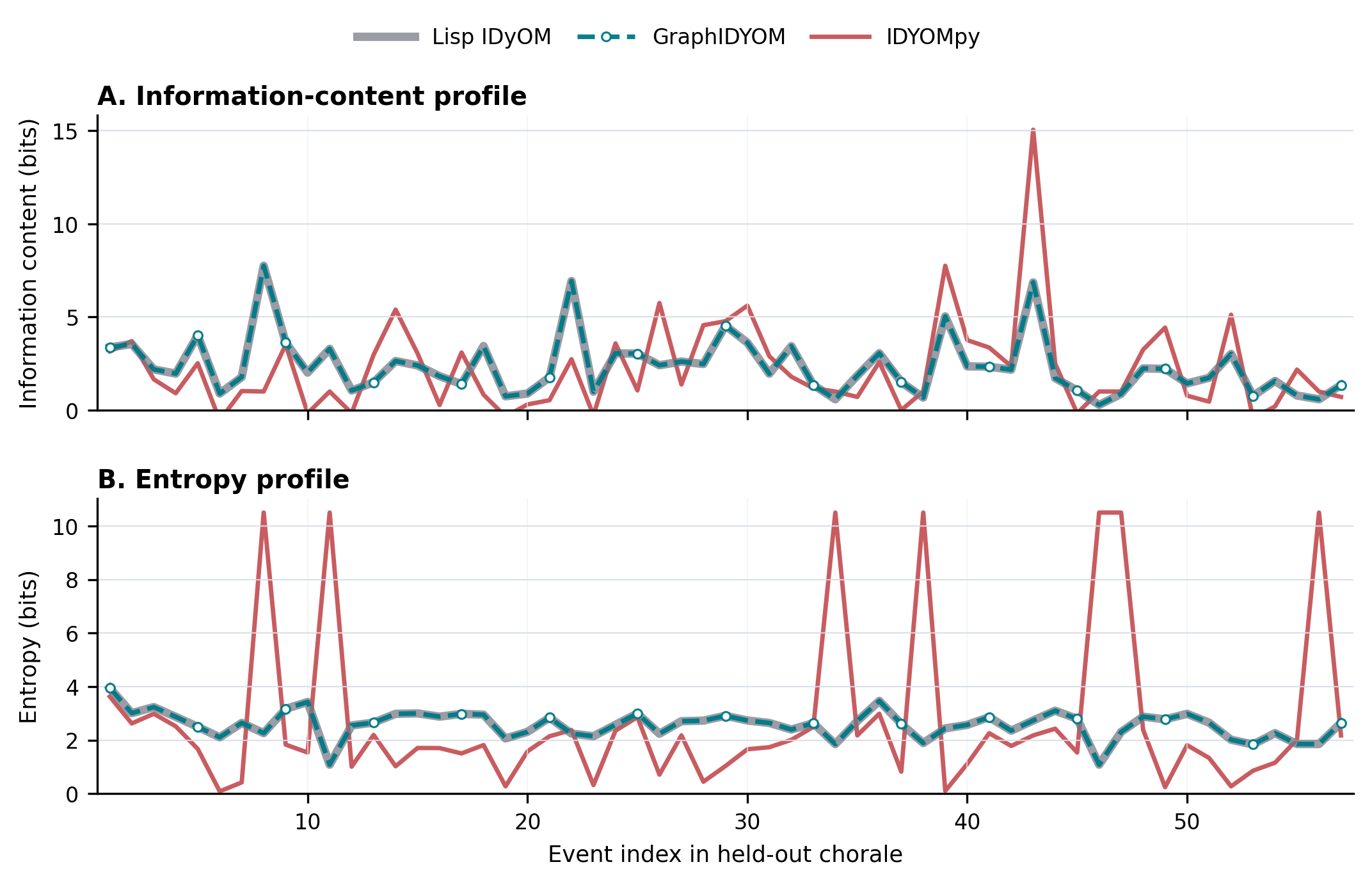}
    \caption[Example event-wise validation profiles]{
    \textbf{Example event-wise validation profiles.}
    Information-content and entropy traces for a held-out Bach chorale
    (\texttt{chor-072}, 57 events) under direct pitch-plus-octave prediction.
    The Lisp reference is plotted as a wide grey trace and GraphIDyOM as a
    dashed teal trace with open markers, making their near-complete overlap
    visible. GraphIDyOM was evaluated under a matched Lisp configuration,
    while IDyOMpy used the closest available corresponding settings.
    }
    \label{fig:validation-profile-example}
\end{figure}

Figure~\ref{fig:validation-profile-example} illustrates the comparison for an
individual held-out chorale under direct pitch-plus-octave prediction. The
GraphIDyOM information-content and entropy traces are visually
indistinguishable from the Lisp reference, whereas IDyOMpy shows larger
deviations under the closest available configuration.

\subsection{Computational performance}

We next compared computational performance using a restricted configuration
that all three systems could express: direct pitch-plus-octave prediction,
corresponding to \texttt{cpitch} in Lisp IDyOM and \texttt{pitch} in IDyOMpy.
The benchmark tested two forms of scaling. First, maximum Markov order was
fixed at 5 while the LTM training set increased from 50 to 100 and 150 files
(2540, 5178, and 7408 training events). Second, the LTM was fixed at 150 files
while maximum order increased from 5 to 10 and 15. The same 35 held-out files
(1819 events) were used for prediction in all conditions.

\begin{figure}[htbp!]
    \centering
    \includegraphics[width=0.7\textwidth]{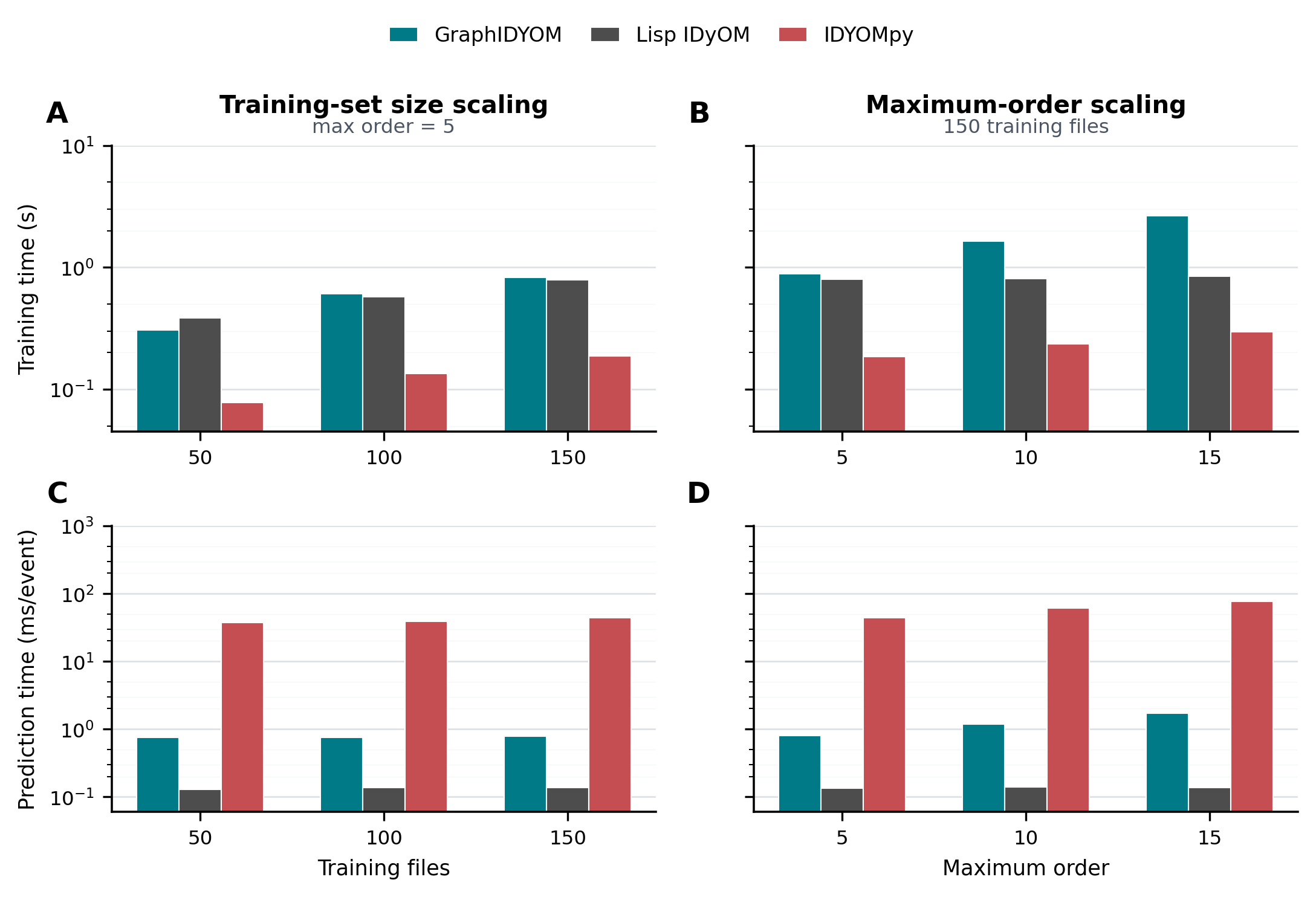}
    \caption[Pitch-octave computational performance benchmark]{
    \textbf{Pitch-octave computational performance benchmark.}
    Computational profile for direct pitch-plus-octave prediction on the
    Bach dataset. The left column varies the number of LTM training files
    while keeping maximum Markov order fixed at 5. The right column fixes
    the LTM at 150 training files and varies maximum order. The top row shows
    LTM construction time; the bottom row shows average held-out prediction
    time per event while the STM is updated online. All y-axes use logarithmic scales.
    }
    \label{fig:pitch-octave-performance}
\end{figure}

Each implementation and condition was evaluated in a fresh worker process. We
measured separately (i) parsing and LTM-construction time and (ii) held-out
prediction time while the STM was updated online.
Figure~\ref{fig:pitch-octave-performance} reports prediction time per event
rather than total batch time, providing a more direct estimate of the cost of
successive prediction requests. The reference Lisp IDyOM runs were invoked
through the Py2LispIDyOM wrapper \citep{guan2022py2lispidyom}; training and
prediction were called separately so that LTM construction and held-out
prediction could be timed independently.

The GraphIDyOM implementation constructed the LTM in less than one second
across the training-size conditions and remained within a few seconds at
maximum order 15. With 150 training files, average prediction time increased
from approximately 0.8~ms per event at order 5 to 1.7~ms at order 15. These
values support low-latency use in interactive symbolic-music applications,
although they are batch-computation averages rather than hard real-time
guarantees.

IDyOMpy showed lower LTM-construction costs but substantially higher prediction
latency in this benchmark, increasing from approximately 44 to 77~ms per event
across the order-scaling conditions. With
\texttt{use\_original\_PPM=True}, its benchmarked prediction path reconstructs
the STM from the current sequence prefix and evaluates probabilities across the
model alphabet. By contrast, the graph-native implementation maintains the STM
incrementally, querying the outgoing continuation counts of the current context
at each order before updating the memory with the newly observed event. This
shifts some cost toward model construction, where explicit NetworkX nodes,
edges, and attributes are materialised, while avoiding repeated reconstruction
during sequential prediction.

The original Lisp implementation remained faster in this minimal direct-call
benchmark. The aim of the Python implementation is therefore not to outperform
the reference system computationally, but to retain low-millisecond prediction
within the same framework used for memory inspection, graph export, model
extension, and external application interfaces. These latencies are compatible
with applications that repeatedly query the model as a symbolic musical
sequence unfolds, providing a practical basis for the interactive and adaptive
uses described in Section~\ref{sec:external_access}.

\section{Extensions enabled by the graph-native framework}
\label{sec:extensions}

The previous sections established GraphIDyOM as a faithful graph-based
reimplementation of IDyOM. The same graph-native representation also makes the
model's internal memories and event-wise computations directly available for
network analysis, controlled modification, and external interaction. These
capabilities extend the validated default predictor without replacing its
underlying prediction architecture.

We demonstrate three such extensions. First, the internal memories and
event-wise outputs can be analysed as network objects, allowing learned
topology, musical structure, and expectation-derived quantities to be studied
within a common representation. 
Second, the effective contribution of stored observations can be modified to instantiate alternative assumptions about
memory retrieval, connecting IDyOM to recency-sensitive PPM models
\citep{harrison2020ppm}. 
Third, trained models can be queried through a local
server, making next-event distributions available to external applications.
These examples are intended as methodological demonstrations of how new
analyses and modelling assumptions can be built around the same validated
IDyOM prediction loop.

\subsection{Network analysis of musical expectation}

Network approaches characterise musical structure through transition topology,
centrality, community organisation, entropy, and related structural measures
\citep{nardelli_tonal_2021,ferretti2017,liu_tse_small_2009,
kulkarni2024information,rossello2025network,di2025decoding}. Predictive models,
by contrast, associate musical events with listener-relative quantities such as
entropy and information content
\citep{sauve_information-theoretic_2019,gold2019predictability,
cheung2019uncertainty,mas-herrero_predictive_2025}. GraphIDyOM makes these
descriptions directly comparable by exposing both predictive memory and
event-wise expectation in graph form.

The graph-native representation therefore shifts the question from
\emph{when} uncertainty or surprise occurs to \emph{where and how} it is
organised over musical topology. It becomes possible to ask whether surprise
is concentrated on central or peripheral states, whether uncertain contexts
correspond to hubs or bridges, or whether unexpected transitions occur between
otherwise well-supported regions of the learned structure.

\subsubsection{Analysing the internal topology of memory}

A direct use of the graph-native implementation is to analyse the topology of
the predictive memories themselves. The LTM can be interpreted as a graph of
accumulated exposure, in which nodes represent viewpoint states or contexts and
weighted edges represent learned continuations. The STM provides the
corresponding online structure accumulated from the current piece. Across
increasing Markov orders, these networks become progressively more
context-specific and typically sparser.

\begin{figure}[htbp!]
    \centering
    \includegraphics[width=0.75\textwidth]{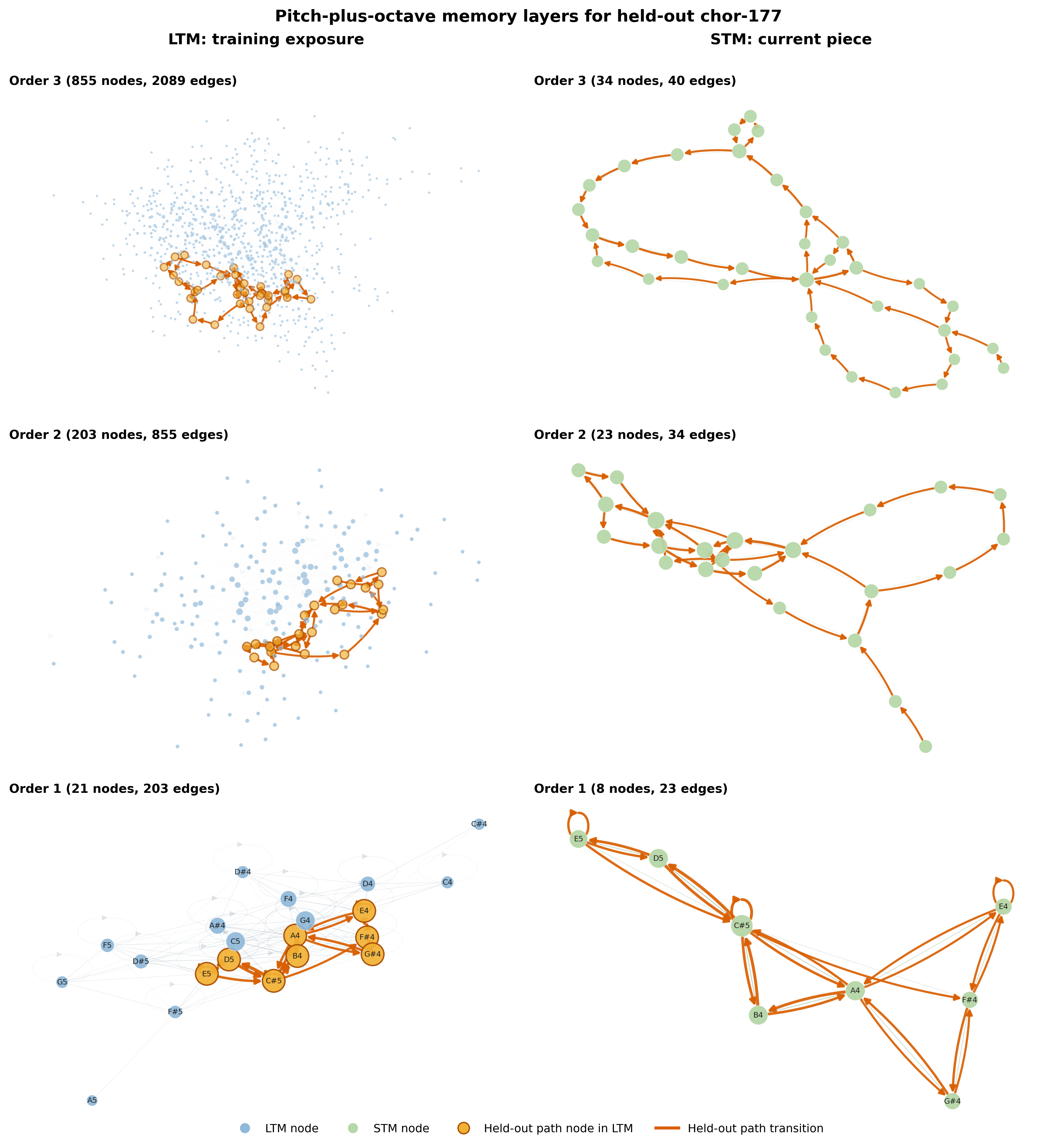}
    \caption[Example network analysis of GraphIDyOM memories]{
    \textbf{Example network analysis of GraphIDyOM memories.}
    Pitch-plus-octave memory graphs for one held-out Bach chorale. The
    left column shows LTM layers trained from 150 other Bach files, and
    the right column shows STM layers accumulated while processing the
    held-out piece. Rows show Markov orders 3, 2, and 1. Orange nodes and edges
    indicate the trajectory of the held-out piece through the corresponding LTM
    layer. All layers are plotted using deterministic force-directed layouts.
    }
    \label{fig:network-analysis-example}
\end{figure}

Figure~\ref{fig:network-analysis-example} illustrates this representation for
the pitch-plus-octave viewpoint. The LTM was trained on 150 Bach files and a
held-out chorale was then processed to construct the STM. LTM and STM are shown
as two columns and Markov orders 3, 2, and 1 as rows. In the LTM layers, the
orange path marks the trajectory of the held-out piece through the learned
memory, while the STM layers show the structure accumulated from the piece
itself.

A melody can therefore be analysed as a trajectory through learned exposure.
Its path may remain within dense, well-supported regions of the LTM, move
through peripheral contexts, or repeatedly require fallback to shorter
histories. These topological properties can be related to the event-wise
entropy and information content produced by the model. GraphIDyOM thus makes
it possible to study predictive uncertainty in relation to \emph{where} a
musical sequence lies within learned memory, rather than only through
piece-level summaries or expectation time series.

\subsubsection{Results as networks of expectations}
\label{sec:expectation_networks}

GraphIDyOM can also project event-wise prediction outputs onto the transition
network of the musical artifact itself. This converts an expectation trace from
a purely temporal representation into annotations over recurring states and
transitions, allowing repeated visits to the same part of the musical topology
to be compared directly.

The structural network describes transitions present in the piece itself.
The expectation-annotated version retains the same topology but associates its
nodes and edges with IDyOM-derived quantities. Node annotations can summarise
predictive entropy when a state is visited, while edge annotations can
summarise the information content of particular transitions. 
Variability across repeated visits can also be analysed, revealing whether the predictive role of a state or transition is stable or changes with higher-order context, short-term adaptation, or position within the piece.

The two descriptions need not agree. A state with several competing continuations may be structurally uncertain
within the piece yet strongly predictable for a model whose LTM contains
well-supported continuations from that state. Conversely, a
locally simple transition may remain surprising if it is weakly supported by
prior exposure. Expectation-annotated networks therefore expose where learned
experience stabilises, amplifies, or departs from the uncertainty implied by
the musical artifact alone.

\begin{figure}[htbp!]
    \centering
    \includegraphics[width=0.7\textwidth]{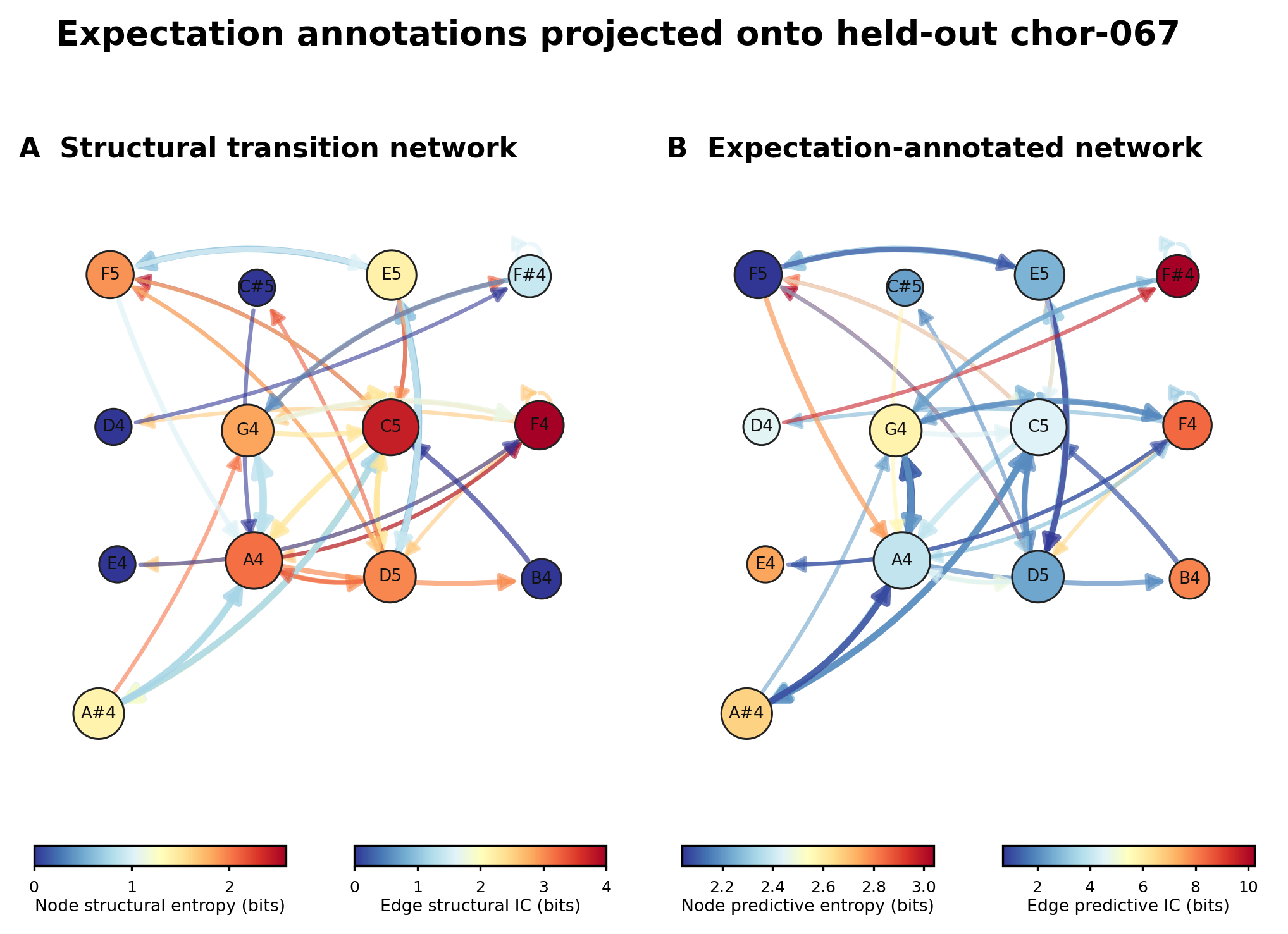}
    \caption[Expectation annotations projected onto a musical network]{
    \textbf{Expectation annotations projected onto a musical network.}
    Example pitch-plus-octave network for the Bach held-out chorale
    (\texttt{chor-067}; 89 events, 13 nodes, 42 edges).
    Panel A shows the structural transition network of the piece: node colour
    gives first-order structural entropy, edge colour gives first-order
    structural information content, node size gives visit count, and edge width
    gives transition count. Panel B uses the same topology and layout, but
    colours nodes by predictive entropy and edges by predictive
    information content from an LTM-only model trained on 150 Bach files
    with maximum order 5. The LTM-only condition isolates expectations derived
    from prior exposure from statistics learned online from the displayed
    piece.
    }
    \label{fig:expectation-network-example}
\end{figure}

Figure~\ref{fig:expectation-network-example} shows this representation for a
held-out Bach chorale. Panel A describes the piece using its own first-order
pitch-plus-octave transition statistics. Panel B retains the same topology but
annotates it with average predictions from an LTM-only model trained on
other Bach chorales. The LTM-only condition is used to isolate expectations
derived from prior exposure from statistics learned online from the displayed
piece.

In this example, the two representations show a clear contrast between
structural and predictive uncertainty. Several of the more central and
frequently visited states have relatively high structural entropy in
Panel~A, reflecting multiple possible continuations within the piece, but
lower predictive entropy in Panel~B, indicating that these continuations are
more strongly constrained by the learned LTM. Peripheral states show the
opposite tendency, appearing structurally simpler within the piece while
remaining comparatively uncertain or surprising to the predictive model.
Although this single example is illustrative rather than generalisable, it
demonstrates how the same musical topology can support different notions of
uncertainty depending on whether they are derived from the artifact itself or
from learned expectations.

\subsection{Recency-sensitive memory retrieval}

IDyOM treats stored context--continuation observations as empirical counts whose
contribution does not decrease with age. Harrison and colleagues extended this
framework with recency-sensitive PPM, in which older observations contribute
progressively less strongly to prediction \citep{harrison2020ppm}. The present
implementation incorporates this mechanism as an optional transformation of
the counts retrieved from the same predictive memory. Conceptually, each stored observation retains the same structural support in
memory but contributes an age-dependent weight when a prediction is made.
Recent observations may retain their full contribution within a configurable
buffer, whereas older evidence is progressively down-weighted according to the
retrieval function. The graph topology is therefore preserved while the
effective strength of its transitions changes with recency. 

\begin{figure}[htbp!]
    \centering
    \includegraphics[width=0.75\textwidth]{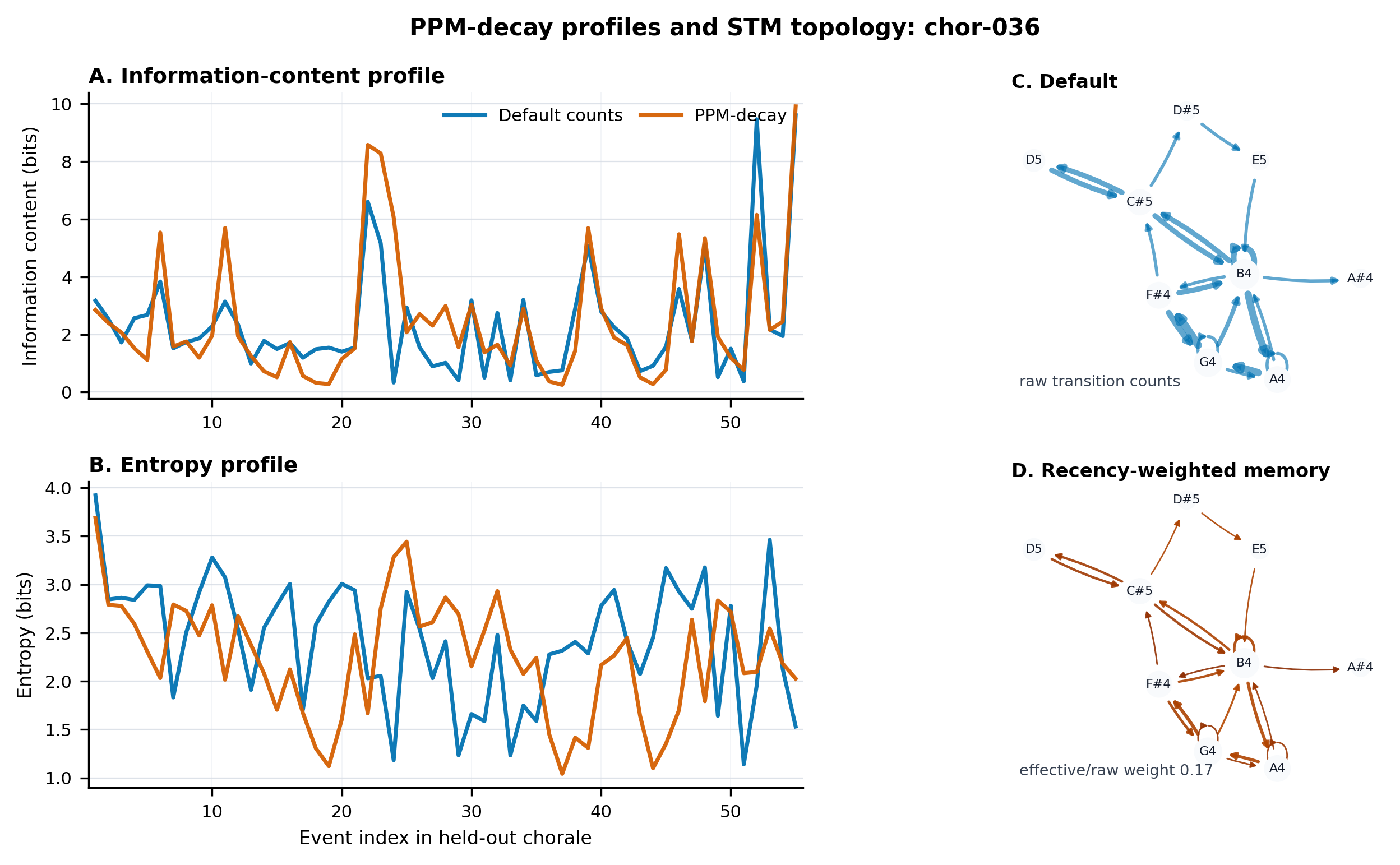}
    \caption[Information-content, entropy, and STM topology under PPM-decay]{
    \textbf{Information-content, entropy, and STM topology under PPM-decay.}
    Event-wise profiles for the held-out Bach chorale
    \texttt{chor-036} (55 events). Panels A and B show information content and
    predictive entropy under default empirical counts and deterministic
    recency-weighted retrieval. Panels C and D show the final order-1 STM
    graph under the same two conditions. The transition support is preserved,
    while edge width and colour in Panel D reflect the effective
    recency-weighted retrieval strengths. Both conditions use the same parser,
    viewpoint, training split, maximum order, and LTM+STM configuration; only
    the optional retrieval mechanism differs.
    }
    \label{fig:imperfect-memory-profiles}
\end{figure}

Figure~\ref{fig:imperfect-memory-profiles} illustrates the effect of
recency-sensitive retrieval on a held-out Bach chorale. Panels A and B show
event-wise predictions from the full LTM+STM model, while Panels C and D show
the final order-1 STM graph accumulated over the same piece. The two conditions
use the same pitch-plus-octave configuration and LTM training set and differ
only in whether stored observations contribute through empirical counts or
deterministic recency-weighted counts. In the recency-sensitive condition,
PPM-decay is applied to both LTM and STM; the network panels visualise only the
final STM, with edge widths and colours indicating effective retrieval weights
at the final query time. Stored observations are preserved, but their
contribution decreases with age once they leave the item buffer, giving greater
weight to more recent evidence. These settings are used only to demonstrate
the representational consequences of the mechanism.

The resulting profiles show that changing retrieval weights affects not only
the event-wise entropy and information content produced by the model, but also
the effective structure of the predictive memory from which those quantities
arise. Because the memory is represented explicitly as a graph, the
consequences of recency-sensitive retrieval can be inspected directly through
changes in edge weights while the stored transition support remains fixed.
This makes it possible to analyse the same memory manipulation at two levels:
in its effect on the prediction profile and in its effect on the internal
representation supporting those predictions.

\subsection{External access and interactive applications}
\label{sec:external_access}

GraphIDyOM can keep a trained model loaded in a local Python process and expose
its predictions through a lightweight JSON server. External applications can
therefore send an unfolding musical history and receive next-event candidates
and their probabilities without embedding the implementation
itself. LTM-only, STM-only, and combined prediction modes are available, while
more detailed entropy, information-content, and order-wise traces remain
accessible through the Python API.

This access can support experimental workflows in which expectation is treated
as a design variable rather than only as a post-hoc descriptor. Studies of
musical expectation commonly construct sets of musical stimuli and subsequently
characterise their uncertainty or surprise
\citep{mas-herrero_predictive_2025,bianco2020long}. With direct access to the
next-event distribution, stimulus construction can instead proceed
iteratively: candidate continuations can be evaluated or selected while the
sequence is being generated, allowing stimuli to be steered toward desired
levels or temporal profiles of entropy and information content. This could
facilitate the construction of controlled contrasts in which predictive
properties are manipulated more directly while other musical constraints are
maintained.

The same mechanism supports interactive musical applications. An external
system can repeatedly query the model as music unfolds and use the returned
probabilities as suggestions, constraints, visual feedback, or inputs to
generation. Expectation can therefore become an active parameter of musical
interaction rather than a quantity computed only after the musical material
has been produced.

\begin{figure}[htbp!]
    \centering
    \includegraphics[width=0.8\textwidth]{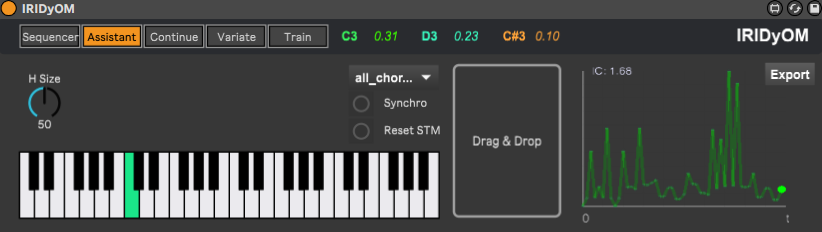}
    \caption[Interactive deployment of GraphIDyOM.]{
    \textbf{Interactive deployment of GraphIDyOM.}
    Assistant mode of the IRIDyOM Max for Live interface integrated in Ableton
    Live. GraphIDyOM provides the predictive backend, returning likely
    next-note candidates and event-wise information-content estimates while the
    musical sequence is inspected or edited. Adapted from
    \citet{bonorossello_iridyom_2026}.
    }
    \label{fig:iridyom-assistant}
\end{figure}

IRIDyOM, a prototype developed alongside GraphIDyOM, demonstrates this
application pathway by integrating the predictive model directly into Ableton
Live through a Max for Live interface
\citep{bonorossello_iridyom_2026}. As illustrated in
Figure~\ref{fig:iridyom-assistant}, the interface can expose the current
predictive state during composition, including likely next-note candidates and
the evolving information-content profile of the musical sequence. The same server also supports controlled variation, continuation, and
expectation-guided generation. Its integration within a widely used digital
audio workstation demonstrates that the prediction framework can operate not
only as an offline analytical tool, but as an accessible component of an
established musical production environment.

\section{Discussion}

GraphIDyOM closely reproduced the original Lisp IDyOM across direct,
projected, and multiple-viewpoint configurations, indicating that the core
prediction mechanisms are consistently preserved under matched symbolic input
and model settings. The model retains native Python
execution while covering a broader subset of the tested reference architecture
than IDyOMpy and showing substantially lower event-wise prediction latency in
the present benchmark \citep{marion2025idyompy}; Py2LispIDyOM, by contrast,
provides Python access while retaining the original Lisp backend
\citep{guan2022py2lispidyom}. GraphIDyOM's contribution is therefore not a new
predictive model, but an accessible and extensible computational basis in
which the mechanisms generating IDyOM expectations are directly exposed.

The main methodological contribution is the explicit representation of
predictive memory as a set of order-specific graphs. This representation does
not alter the count-based logic of IDyOM, but changes what can be observed and
manipulated within the model. Long-term exposure, short-term adaptation, the
contexts supporting a prediction, and the trajectory followed by a target
sequence can all be inspected within the same representation. The model is
therefore no longer available only through its final probabilities or
information-theoretic outputs: the internal memory structures from which those
outputs arise become objects of analysis in their own right.

The recency-sensitive memory example illustrates this advantage directly.
PPM-decay preserves which observations have been stored while changing their
effective contribution as they become older \citep{harrison2020ppm}. In
the graph-native implementation, the consequences of this manipulation can be observed not only in
the resulting entropy and information-content profiles, but also within the
memory representation itself through changes in effective edge weights. The
same intervention can therefore be examined at two linked levels: how a memory
assumption changes the internal state of the model and how those changes
propagate to prediction. This creates a common framework for testing alternative
assumptions about retrieval, forgetting, or other transformations of learned
evidence without replacing the surrounding prediction architecture.

The graph-based connection between musical structure and predictive output
provides a complementary methodological opportunity. Statistical, geometrical,
and network descriptions primarily characterise organisation present in the
musical artifact, whereas IDyOM estimates quantities that depend on the
regularities acquired by a particular model through prior exposure and local
context. These two descriptions may interact without coinciding. A state that
is structurally highly connected or uncertain within a piece may nevertheless
be strongly predicted from prior exposure, while a locally simple transition
may remain unexpected for a model in which it is weakly represented. This
distinction provides one route for studying how artifact-level complexity and
listener-relative or perceived complexity relate to one another
\citep{sauve_information-theoretic_2019}.

Mapping predictive entropy and information content onto recurring musical
states and transitions makes this relation directly analysable. Two pieces, or
two listener models, may have similar average information-theoretic values
while organising those values very differently over their topology: surprise
may be concentrated on peripheral transitions, distributed across central
states, or associated with movement between otherwise stable regions. Moreover,
a state or transition need not have a single predictive role. Repeated visits
can carry different values depending on higher-order context, prior exposure,
and short-term adaptation, and the model retains these visit-level quantities
rather than restricting the representation to their averages. Such analyses
may be useful for work relating musical expectation to perceived complexity,
pleasure, and preference
\citep{sauve_information-theoretic_2019,gold2019predictability,
bianco2019music,cheung2019uncertainty,mas-herrero_predictive_2025,
berlyne1971aesthetics}, although dedicated perceptual studies would be required
to establish specific relationships between network organisation and
subjective experience.

External access adds a third dimension by making expectation available during,
rather than only after, the construction of musical material. In experimental
work, musical stimuli are often created and subsequently characterised with
IDyOM-derived uncertainty or surprise. A queryable predictive model instead
allows candidate events or sequences to be evaluated while stimuli are being
designed, making it possible to steer them toward particular levels or temporal
profiles of entropy and information content. This could support more direct
manipulation of predictive variables in studies of musical expectation,
preference, or neural response.

The same mechanism allows GraphIDyOM to participate directly in musical
interaction. The IRIDyOM prototype demonstrates this possibility by integrating
the model within Ableton Live, where predictions can be inspected and used for
expectation-guided continuation, variation, and generation
\citep{bonorossello_iridyom_2026}. Such an interface could also provide a
practical route for extending expectation modelling beyond the currently
implemented symbolic dimensions. Additional musical attributes---for example
dynamics, instrumentation, timbral categories, or other production-related
features---could be encoded as viewpoints and exposed alongside melodic and
temporal expectations, allowing their interactions to be studied within the
same predictive framework. Because predictions and interaction histories can
be retained over time, the same infrastructure could also support analyses of
creative processes themselves \cite{boden2004creative,wiggins2006preliminary,daikoku2021statistical}, relating successive compositional decisions to
the expectations available when those decisions were made. More broadly,
independently initialised models with different stylistic priors or memory
assumptions could provide a basis for future work on distributed musical
expectations and social creativity \cite{saunders_computational_2015,linkola2019creative,bono2024emergent}.

Several limitations remain. First, GraphIDyOM does not yet reproduce the full
viewpoint ecosystem of Lisp IDyOM. The present implementation covers the pitch,
interval, contour, duration, temporal, projection, and multiple-viewpoint
configurations required for the validations reported here, but additional
metrical, harmonic, dynamic, and specialised viewpoints remain to be added.
The strong agreement obtained across the implemented configurations suggests
that extending this coverage should primarily require additional viewpoint
encoding and projection logic rather than changes to the validated prediction
core. Second, the current framework does not reproduce the original automated
viewpoint-selection procedures, including hill-climbing approaches for
identifying effective representations for a dataset. 
The network and recency-sensitive examples presented here are methodological
demonstrations rather than tests of specific perceptual or memory hypotheses.
Validation and performance were also evaluated on selected symbolic datasets
and configurations. Broader testing across repertoires, richer viewpoint combinations, and deployed interactive
environments will therefore be needed to characterise the limits of numerical
correspondence, computational performance, and the analytical possibilities of
the framework more fully.

Taken together, these results suggest that the main value of GraphIDyOM lies
in making the internal structure of musical expectation accessible without
discarding the established IDyOM framework. IDyOM provides a way to quantify
expectation through probability, entropy, and information content; GraphIDyOM
adds explicit access to the learned memories from which those quantities arise
and makes the resulting model directly queryable from external applications.
This makes it possible to inspect how prior exposure and local adaptation shape
prediction, modify assumptions about memory retrieval, relate
listener-relative expectation to the structure of the musical artifact, and
use predictive quantities during experimental or creative interaction.
GraphIDyOM therefore extends IDyOM from a source of predictive outputs into an
inspectable, modifiable, and interactive framework for studying how musical
expectations are represented, updated, organised, and used.

\section*{Code availability}

The GraphIDyOM framework, including source code, documentation, examples, and
tests, is available at
\url{https://github.com/llucbono/GraphIDyOM}. 

\section*{Acknowledgments}

The author acknowledges support from the ARIAC project (No. 2010235), funded by the Service public de Wallonie (SPW Recherche).

\bibliographystyle{unsrtnat}
\bibliography{refs}

\end{document}